\documentclass[12pt, draftcls, onecolumn]{IEEEtran}
\usepackage{setspace}
\usepackage{amsmath,amsfonts}
\usepackage{algorithmic}
\usepackage{algorithm}
\usepackage{array}
\usepackage[caption=false,font=normalsize,labelfont=footnotesize,textfont=sf]{subfig}
\usepackage{textcomp}
\usepackage{stfloats}
\usepackage{url}
\usepackage{color}
\usepackage{verbatim}
\usepackage{graphicx}
\usepackage{cite}
\begin{document}

\title{\Large A Tutorial on Holographic MIMO Communications—Part II: Performance Analysis and Holographic Beamforming}
\author{Jiancheng An,~\IEEEmembership{Member,~IEEE}, Chau Yuen,~\IEEEmembership{Fellow,~IEEE},\\Chongwen Huang,~\IEEEmembership{Member,~IEEE}, M\'erouane Debbah,~\IEEEmembership{Fellow,~IEEE},\\H. Vincent Poor,~\IEEEmembership{Life Fellow,~IEEE}, and Lajos Hanzo,~\IEEEmembership{Life Fellow,~IEEE}\\
\emph{(Invited Paper)}
\thanks{This research is supported by the Ministry of Education, Singapore, under its MOE Tier 2 (Award number MOE-T2EP50220-0019). This work was supported by the Science and Engineering Research Council of A*STAR (Agency for Science, Technology and Research) Singapore, under Grant No. M22L1b0110. The work of Prof. Huang was supported by the China National Key R\&D Program under Grant 2021YFA1000500, National Natural Science Foundation of China under Grant 62101492, Zhejiang Provincial Natural Science Foundation of China under Grant LR22F010002, Zhejiang University Global Partnership Fund, Zhejiang University Education Foundation Qizhen Scholar Foundation, and Fundamental Research Funds for the Central Universities under Grant 2021FZZX001-21. H. V. Poor would like to acknowledge the financial support of the U.S. National Science Foundation under Grant CNS-2128448. L. Hanzo would like to acknowledge the financial support of the Engineering and Physical Sciences Research Council projects EP/W016605/1 and EP/X01228X/1 as well as of the European Research Council's Advanced Fellow Grant QuantCom (Grant No. 789028). \emph{(Corresponding author: Chau Yuen.)}}
\thanks{J. An is with the Engineering Product Development Pillar, Singapore University of Technology and Design, Singapore 487372 (e-mail: jiancheng\_an@sutd.edu.sg). C. Yuen is with the School of Electrical and Electronics Engineering, Nanyang Technological University, Singapore 639798 (e-mail: chau.yuen@ntu.edu.sg). C. Huang is with College of Information Science and Electronic Engineering, Zhejiang-Singapore Innovation and AI Joint Research Lab and Zhejiang Provincial Key Laboratory of Info. Proc., Commun. \& Netw. (IPCAN), Zhejiang University, Hangzhou 310027, China. (e-mail: chongwenhuang@zju.edu.cn ). M. Debbah is with Khalifa University of Science and Technology, P O Box 127788, Abu Dhabi, UAE (email: merouane.debbah@ku.ac.ae). H. Vincent Poor is with the Department of Electrical and Computer Engineering, Princeton University, Princeton, NJ 08544 USA (e-mail: poor@princeton.edu). L. Hanzo is with the School of Electronics and Computer Science, University of Southampton, SO17 1BJ Southampton, U.K. (e-mail: lh@ecs.soton.ac.uk).}}
\maketitle

\begin{abstract}
As Part II of a three-part tutorial on holographic multiple-input multiple-output (HMIMO), this Letter focuses on the state-of-the-art in performance analysis and on holographic beamforming for HMIMO communications. We commence by discussing the spatial degrees of freedom (DoF) and ergodic capacity of a point-to-point HMIMO system, based on the channel model presented in Part I. Additionally, we also consider the sum-rate analysis of multi-user HMIMO systems. Moreover, we review the recent progress in holographic beamforming techniques developed for various HMIMO scenarios. Finally, we evaluate both the spatial DoF and the channel capacity through numerical simulations.
\end{abstract}

\begin{IEEEkeywords}
Holographic MIMO communications, performance analysis, holographic beamforming, near-field communications.
\end{IEEEkeywords}

\section{Introduction}
\IEEEPARstart{T}{o} meet ever more stringent demands in data rate, connectivity, and reliability, next-generation wireless networks relying on new multiple-input multiple-output (MIMO) technologies known as holographic MIMO (HMIMO) \cite{TWC_2022_Pizzo_Fourier}, or large intelligent surfaces \cite{TSP_2018_Hu_Beyond} are conceived. In contrast to the conventional discrete-element based antenna arrays, HMIMO schemes have a nearly continuous antenna surface that can generate any current distribution to fully exploit the propagation characteristics of an electromagnetic (EM) channel. The spatially continuous aperture of HMIMO schemes enhances the spatial resolution and beamforming gains, which is achieved by harnessing compact radiating elements \cite{arXiv_2023_An_Stacked, CM_2021_Dardari_Holographic}. Furthermore, the electromagnetically large aperture of HMIMO schemes enables near-field communications, which introduces an extra distance dimension to assist communications. When the transceiver aperture is sufficiently large, the number of communication modes depends only on the area of the smallest surface normalized with respect to the squared wavelength, providing extremely large spatial multiplexing gains \cite{JSAC_2020_Dardari_Communicating, arXiv_2022_Renzo_LoS}.

To fully unlock the potential of HMIMO communications, it is crucial to design efficient beamforming techniques based on the newly established HMIMO channel models. Conventional beamforming designs, like zero-forcing (ZF), are generally complex. As such, it is imperative to develop \emph{low-complexity} beamforming designs for practical HMIMO communications \cite{WCL_2023_Jia_Environment, TCOM_2022_An_Low}. Additionally, near-field communications have the unique ability to determine both the angle and the distance of a user. This makes it possible to focus the energy radiated onto a specific location in the Fresnel zone, thereby enhancing the so-called near-field \emph{beam-focusing}\footnote{The terminology of beam-focusing is used in near-field communications for distinguishing it from classic far-field beamforming.} \cite{TWC_2022_Zhang_Beam}. To take advantage of this benefit, effective \emph{distance-aware} beam-focusing schemes must be developed.

Let us now critically appraise the recent advances in HMIMO communications, with a particular focus on their performance analysis and holographic beamforming. The rest of the letter is organized as follows. In Section \ref{sec2}, we discuss the statistics of new HMIMO channel models, which are used for characterizing the spatial degrees-of-freedom (DoF) and HMIMO channel capacity. In Section \ref{sec3}, we briefly review the recent progress in holographic beamforming, while Section \ref{sec4} provides numerical results for illustrating our analysis. Finally, Section \ref{sec5} concludes this letter.
\section{Performance Analysis of HMIMO Communications}\label{sec2}
\subsection{Spatial DoF of HMIMO}
The spatial DoF refers to the number of independent data streams that can be simultaneously transmitted through the wireless propagation environment. When the transceiver aperture is sufficiently large and includes a massive number of elements, the normalized spatial DoF is asymptotically limited by \cite{TSP_2018_Hu_Beyond}
\begin{align}
\text{DoF}=\begin{cases}
 2/\lambda, & \text{{/m}, Linear array,}\\
 \pi/\lambda ^{2}, & \text{{/m$^{2}$}, Planar array,}
 \end{cases}
\label{eq1}
\end{align}
where $\lambda$ is the wavelength. Eq. \eqref{eq1} implies that there is a maximum number of parallel subchannels that can be established on an HMIMO communication link, and this limit can be approached by increasing the size and element density of the transceiver aperture. Expanding planar arrays to volumetric arrays results in only a two-fold increase of the DoF \cite{TWC_2022_Pizzo_Fourier}.

Based on \eqref{eq1}, the DoF loss incurred by disregarding evanescent waves is
\begin{align}
 1-\frac{\pi /\lambda ^{2}}{\left ( 2/\lambda \right )^{2}}=1-\frac{\pi }{4}\approx 21.5\%.
\end{align}

\begin{figure*}[!t]
\centering
\includegraphics[width=16cm]{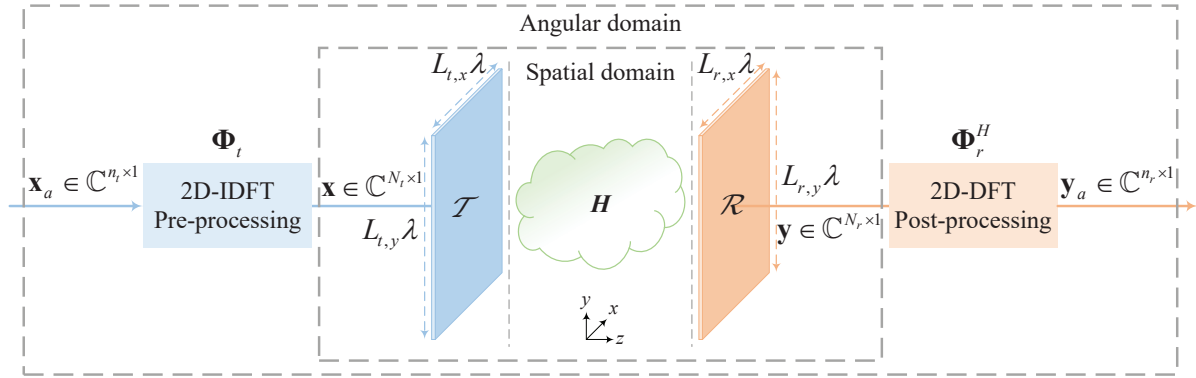}
\caption{A transceiver architecture for HMIMO communications.}
\label{fig_1}
\end{figure*}
Moreover, Dardari \cite{JSAC_2020_Dardari_Communicating} examined the DoF of a point-to-point HMIMO system, relying on arbitrary transceiver configurations and near-field communications. It was revealed that the spatial DoF attained may exceed one even under strong line-of-sight (LoS) channel conditions, leading to a beneficial increase in the HMIMO channel capacity.
\subsection{Power Scaling Law of HMIMO}
Recently, Lu and Zeng \cite{TWC_2022_Lu_Communicating} have derived a closed-form expression for the signal-to-noise ratio (SNR) using optimal single-user maximum ratio transmission (MRT) based beamforming. Their approach takes into account the variations of signal phase, amplitude, and projected aperture across the array elements that are applicable for both far-field and radiative near-field scenarios. It was revealed that the SNR increases with the number of elements, but with diminishing returns. To refine the near- and far-field separation, a uniform-power distance criterion has been proposed, which takes power variations into account. In addition, Wang \emph{et al.} \cite{ICC_2022_Wang_Performances} have examined the system capacity of an LoS-dominated point-to-point HMIMO system. They analytically demonstrated that an LoS-dominated HMIMO system can achieve a power gain of approximately $9.94$ dB in a far-field scenario (corresponding to a SE increase of $3.30$ bits/s/Hz) compared to a conventional MIMO system with half-wavelength spaced isotropic antenna arrays in the same transceiver apertures. As the aperture increases, the signal attenuation and polarization mismatch also increases, leading to a maximum channel power gain of $1/3$ with respect to an isotropic antenna in the near field \cite{CM_2021_Dardari_Holographic}.
\subsection{Channel Capacity Analysis of Point-to-Point HMIMO Systems}
To elaborate, we consider an HMIMO transceiver architecture, as illustrated in Fig. \ref{fig_1}. This architecture involves two parallel, vertically oriented planar arrays that span the rectangular regions $\mathcal{T}$ and $\mathcal{R}$, each having dimensions of $\left (L_{t,x}\lambda \times L_{t,y}\lambda \right )$ and $\left (L_{r,x}\lambda \times L_{r,y}\lambda \right )$, respectively. The transmit array is equipped with $N_{t}$ antennas, while the receive array has $N_{r}$ antennas. Let $\mathbf{H}\in \mathbb{C}^{N_{r}\times N_{t}}$ denote the HMIMO channel having a spatial correlation matrix $\mathbf{R} = \mathbb{E}\left \{ \textrm{vec}\left ( \mathbf{H} \right )\textrm{vec}\left ( \mathbf{H} \right )^{H} \right \}\in \mathbb{C}^{N_{r}N_{t}\times N_{r}N_{t}}$. If the separability condition is imposed, then $\mathbf{R}$ can be rewritten as \cite{TWC_2022_Pizzo_Fourier}
\begin{align}\label{eq3}
 \mathbf{R}=\underbrace{\mathbf{\Phi} _{t}\left [ \textrm{diag}\left ( \boldsymbol{\sigma }_{t} \right ) \right ]^{2}\mathbf{\Phi} _{t}^{H}}_{\mathbf{R}_{t}\in \mathbb{C}^{N_{t}\times N_{t}}}\otimes \underbrace{\mathbf{\Phi} _{r}\left [ \textrm{diag}\left ( \boldsymbol{\sigma }_{r} \right ) \right ]^{2}\mathbf{\Phi} _{r}^{H}}_{\mathbf{R}_{r}\in \mathbb{C}^{N_{r}\times N_{r}}},
\end{align}
where the matrices $\mathbf{\Phi }_{t}\in \mathbb{C}^{N_{t}\times n_{t}}$ and $\mathbf{\Phi }_{r}\in \mathbb{C}^{N_{r}\times n_{r}}$ depend solely on the array layout, while $\boldsymbol{\sigma }_{t}$ and $\boldsymbol{\sigma }_{r}$ describe the power transfer at transmitter and receiver, respectively. Furthermore, the eigenvalues are limited to $n_{r}n_{t}$ by physical principles, with $n_{t}=\left \lceil \pi L_{t,x}L_{t,y} \right \rceil$ and $n_{r}=\left \lceil \pi L_{r,x}L_{r,y} \right \rceil$.

Next, we use the channel model developed in Part I to evaluate the capacity of the HMIMO communication system. To simplify the analysis, we will analyze the HMIMO capacity in the angular domain. Specifically, let $ \mathbf{H}_{a} \in \mathbb{C}^{n_{r}\times n_{t}}$ represent the equivalent HMIMO channel in the angular domain, which has a rank of $r = \textrm{rank}\left ( \mathbf{H}_{a} \right )$ that determines the DoF of the HMIMO channel. As illustrated in Fig. \ref{fig_1}, the transformation from the angular domain to the spatial domain is entirely determined by the semi-unitary matrices $\mathbf{\Phi }_{t}$ and $\mathbf{\Phi }_{r}^{H}$. With the aid of uniform sampling, $\mathbf{\Phi }_{t}$ and $\mathbf{\Phi }_{r}^{H}$ become two-dimensional (2D) discrete Fourier transform (DFT) and 2D inverse DFT (IDFT) matrices, which can be efficiently implemented in the analog domain with the aid of radio-frequency (RF) signal processing, for instance, through the use of lens antenna arrays. Moreover, we assume that $\mathbf{Q}_{a}=\mathbb{E}\left \{ \mathbf{x}_{a}\mathbf{x}_{a}^{H} \right \}\in \mathbb{C}^{n_{t}\times n_{t}}$ with $\textrm{tr}\left ( \mathbf{Q}_{a} \right )\leq 1$. As a result, the ergodic capacity of a point-to-point HMIMO system, measured in bit/s/Hz, is given by
\begin{align}\label{eq6}
 C=\underset{\mathbf{Q}_{a}:\textrm{tr}\left ( \mathbf{Q}_{a} \right )\leq 1}{\max}\mathbb{E}\left \{ \log_{2}\det\left ( \mathbf{I}_{n_{r}} +\gamma \mathbf{H}_{a}\mathbf{Q}_{a}\mathbf{H}_{a}^{H} \right ) \right \},
\end{align}
where $\gamma $ is the SNR at the receiver accounting for the large-scale fading. Next, we will proceed to evaluate the ergodic capacity in \eqref{eq6} for different amounts of channel state information (CSI).
\subsubsection{Statistical CSI at the Transmitter (CSIT)}
Assuming that perfect knowledge of the strength of coupling coefficients is available at the transmitter, the optimal angular covariance matrix is given by $\mathbf{Q}_{a}=\mathbf{P}_{a}$, where $\mathbf{P}_{a}\in \mathbb{C}^{n_{t}\times n_{t}}$ is a diagonal matrix with its diagonal entries representing the optimal powers. The power allocation is determined using the classic water-filling algorithm \cite{TCOM_2022_An_Low}. The optimal $\mathbf{x}_{a}$ is $\mathbf{x}_{a}=\mathbf{P}_{a}^{1/2}\mathbf{s}_{a}$ with $\mathbf{s}_{a}\sim \mathcal{CN}\left ( \mathbf{0},\mathbf{I}_{n_{t}} \right )$ representing the information-bearing vector. This implies that, with known CSIT, the capacity-achieving transmission strategy is to transmit a statistically independent information stream in the angular domain.
\subsubsection{Instantaneous CSI at the Receiver (CSIR)}
With instantaneous CSIR, the ergodic capacity in \eqref{eq6} is achieved by utilizing an input vector $\mathbf{x}_{a}$ consisting of independent and identically distributed (i.i.d.) values, with uniform power allocation among $n_{t}$ parallel channels, i.e., $\mathbf{Q}_{a}=\frac{1}{n_{t}}\mathbf{I}_{n_{t}}$. By doing so, the resultant channel capacity is given by
\begin{align}\label{eq7}
 C=\sum_{i=1}^{r}\mathbb{E}\left \{ \log_{2}\left ( 1+\frac{\gamma }{n_{t}} \lambda _{i}\left ( \mathbf{H}_{a}\mathbf{H}_{a}^{H} \right ) \right ) \right \},
\end{align}
where $ \lambda _{i}\left ( \mathbf{A} \right ) $ represents the eigenvalues of an arbitrary matrix $\mathbf{A}$. For HMIMO channels having separable correlations and electromagnetically large transceiver arrays, \eqref{eq7} can be asymptotically approximated by \cite{TWC_2022_Pizzo_Fourier}
\begin{align}\label{eq8}
 C\approx \sum_{j=1}^{n_{t}}\log_{2}\left ( \frac{1+\gamma \sigma _{t,j}^{2}\Gamma _{r}}{e^{\gamma \Gamma _{t}\Gamma _{r}}} \right )+\sum_{i=1}^{n_{r}}\log_{2}\left ( 1+\gamma\sigma _{r,i}^{2}\Gamma _{t} \right ),
\end{align}
where the coefficients $\Gamma _{t}$ and $\Gamma _{r}$ are determined by iteratively solving the pair of fixed-point equations (60) and (61) in \cite{TWC_2022_Pizzo_Fourier}.
\subsubsection{Perfect CSIT and CSIR}
Assuming that both transceivers have perfect knowledge of $\mathbf{H}_{a}$ and letting $\mathbf{H}_{a}=\mathbf{U}_{a}\mathbf{\Lambda }_{a}\mathbf{V}_{a}^{H}$ denote its singular value decomposition (SVD), the capacity defined in \eqref{eq6} can be achieved by utilizing $\mathbf{x}_{a}=\mathbf{V}_{a}\mathbf{P}_{a}^{1/2}\mathbf{s}_{a}$, where $\mathbf{s}_{a}\in \mathbb{C}^{n_{t}}$ is an i.i.d. circularly symmetric complex Gaussian (CSCG) vector with unit variance. As a result, the covariance matrix of $\mathbf{x}_{a}$ is given by $\mathbf{Q}_{a}=\mathbf{V}_{a}\mathbf{P}_{a}\mathbf{V}_{a}^{H}$, and the resultant ergodic capacity is formulated as \cite{TWC_2022_Pizzo_Fourier}
\begin{align}\label{eq9}
 C=\sum\limits_{i=1}^{ r }\log_{2}\left [ \mu \lambda _{i}\left ( \mathbf{H}_{a}\mathbf{H}_{a}^{H} \right ) \right ]^{+},
\end{align}
where $\mu$ is the water-filling level determined such that $\gamma =\sum\nolimits_{i=1}^{r}\left [ \mu -1/\lambda _{i}\left ( \mathbf{H}_{a}\mathbf{H}_{a}^{H} \right ) \right ]^{+}$.

\subsection{Sum-Rate Analysis of Multiuser HMIMO Systems}
Furthermore, there are significant concerns regarding the performance analysis of multiuser HMIMO communications. For instance, Yuan \emph{et al.} \cite{TCOM_2020_Yuan_Towards} investigated the uplink performance of large intelligent surfaces (LIS)-based communication using matched filtering (MF) for the far-field scenario. The study theoretically characterized the array gain, spatial resolution, and capability of interference suppression concerning various factors, such as the LIS size, orientation, and geographic deployment. Specifically, as the array aperture of a centralized LIS increases, the overall sum rate of $K$ users can be approximated as
\begin{align}
 R\approx \sum\nolimits_{k=1}^{K}\log_{2}\left ( 1+\frac{p_{k} \varepsilon _{k}}{\sigma ^{2}}\pi r_{\text{LIS}}^{2} \right ),
\end{align}
where $p_k$ is the transmitted power at the $k$-th user, $\varepsilon _{k}$ is the free-space path loss between the transmitter and the antenna of the $k$-th user, while $\sigma ^{2}$ and $r_{\text{LIS}}$ represent the average noise power and the radius of the circular LIS, respectively. Besides, an appropriate user association scheme, orientation control, and power control are capable of expanding the coverage area of a distributed LIS system. Moreover, Jung \emph{et al.} \cite{TWC_2020_Jung_Performance} conducted an asymptotic analysis of the uplink data rate in an HMIMO system taking into consideration the channel estimation errors and hardware impairments. The simulation results showed that as the number of antennas increases, the impact of hardware impairments, noise, channel estimation errors, and non-line-of-sight (NLoS) interference becomes less grave. The study also revealed that a LIS can provide a comparable rate to massive MIMO with improved reliability and reduced antenna area.

\section{Holographic Beamforming}\label{sec3}
In this section, we review recent holographic beamforming schemes. Specifically, we focus on four typical scenarios, which are \emph{i)} point-to-point, \emph{ii)} multi-user, \emph{iii)} wideband and \emph{iv)} near-field HMIMO scenarios.
\subsection{Point-to-Point HMIMO Scenarios}
To enable point-to-point HMIMO communications in practice, Sanguinetti \emph{et al.} \cite{TWC_2022_Sanguinetti_Wavenumber} proposed a wavenumber-division multiplexing (WDM) scheme. Specifically, they first developed a channel model, where the continuous transmit currents and received fields are represented by using Fourier basis functions. The orthogonality among different communication modes is guaranteed as the receiver size tends to become infinitely large. By operating in the wavenumber domain, the WDM scheme provides a low-complexity signal processing architecture to mitigate the interference, while providing the same SE as the optimal singular-value decomposition (SVD) aided transmission. Nonetheless, the WDM scheme proposed in \cite{TWC_2022_Sanguinetti_Wavenumber} only accounts for HMIMO communications between two parallel line segments that are in LoS conditions. In practical scenarios, however, the source and receiver may be 2D surfaces that can have arbitrary positions and orientations and there might be other factors such as multipath propagation and LoS blockage. Hence, it is necessary to conduct further investigations to fully exploit the WDM scheme under practical conditions.
\subsection{Multiuser HMIMO Scenarios}
Employing multiuser transmit beamforming is crucial for mitigating the co-channel interference experienced by the downlink receivers \cite{TGCN_2022_An_Joint}. In \cite{TWC_2022_Deng_Reconfigurable}, Deng \emph{et al.} explored the use of reconfigurable holographic surfaces (RHS) in the downlink of a multi-user HMIMO system. They showed that the RHS can accurately steer multiple beams with low sidelobe levels (less than $-8$ dB) using amplitude-controlled holographic beamforming. Additionally, the RHS was found to outperform the identical-sized phased array in terms of sum rate, while also reducing the hardware costs in practical applications. Furthermore, You \emph{et al.} \cite{TWC_2023_You_Energy} studied the uplink of dynamic metasurface antenna (DMA)-assisted multi-user HMIMO systems. They performed joint optimization of the transmit precoding (TPC) matrices and DMA tuning weights at the base station (BS) for maximizing the energy efficiency (EE). The results indicated that DMA-assisted HMIMO communications are capable of achieving higher EE than conventional large-scale antenna arrays, particularly in high-power budget situations. The impact of imperfect CSI on the performance of DMA-assisted multi-user HMIMO systems was also evaluated. Note that designing robust HMIMO transmission schemes that account for imperfect CSI will be of practical interest. Additionally, Zhang \cite{arXiv_2021_Zhang_Continuous} developed a pattern-division multiplexing (PDM) technique for multi-user HMIMO systems. This approach transformed the design of continuous pattern functions into the design of their projection lengths on finite orthogonal bases, thereby overcoming the challenge of functional programming. A block coordinate descent (BCD) based scheme was proposed for solving the sum-rate maximization problem formulated. Simulation results showed that PDM achieved a three-fold higher sum-rate than the WDM scheme.

\subsection{Wideband HMIMO Scenarios}
In wideband HMIMO systems, effective holographic beamforming schemes are urgently needed for addressing deleterious beam squint effects. In \cite{TCOM_2021_Wang_Dynamic}, Wang \emph{et al.} explored the application of DMAs in orthogonal frequency division multiplexing (OFDM) receivers equipped with low-resolution analog-to-digital converters (ADCs). Due to their intrinsic structure, DMAs possess a controllable frequency-selective profile, thus they are capable of providing frequency-domain diversity that enhances the performance at a lower cost and power consumption than conventional hybrid architectures. Additionally, Xu \emph{et al.} \cite{arXiv_2022_Xu_Near} proposed a beam-combining scheme for near-field wideband holographic metasurface antenna (HMA)-assisted HMIMO uplink transmissions, based on a spherical wavefront model that accounts for both the near-field and dual-domain selectivity. An alternating optimization method was employed for beam-combining at a low complexity. Moreover, the minorization-maximization approach was used for addressing the non-convex feasible entries of the HMA weight matrix caused by the associated physical structure. The HMA-based HMIMO scheme provides a higher sum rate than its conventional hybrid analog and digital counterpart within the same array aperture. In summary, significant performance improvements can be achieved at a reduced cost and power consumption by exploiting the frequency-domain diversity offered by metasurfaces and by employing advanced holographic beamforming methods.
\subsection{Near-Field HMIMO Scenarios}
When the antenna aperture dimension is comparable to the link distance, the operating conditions fall within the Fresnel region \cite{JSAC_2020_Dardari_Communicating}, where radiating near-field propagation occurs. The operation below the Fraunhofer distance opens up new opportunities for enhancing the performance. By exploiting the distance information and phase variants, effective beam-focusing techniques can achieve favorable performance gains in the near-field region \cite{arXiv_2021_Cui_Near}. For example, Zhang \emph{et al.} \cite{TWC_2022_Zhang_Beam} explored the potential of beam-focusing for multi-user downlink HMIMO scenarios in the near-field region. A mathematical model is first provided to characterize the near-field channels and transmission patterns. The customized focused beams provide a new DoF for mitigating the interference in both the angle and distance domains, which is not achievable using conventional far-field beamforming. Moreover, Ji \emph{et al.} \cite{WCL_2023_Ji_Extra} derived the theoretical expressions of both the DoF and capacity gain in the reactive near-field region, based on the Fourier plane-wave series representation of an HMIMO channel. It is demonstrated that evanescent waves can significantly enhance the spatial DoF and increase capacity in the reactive near-field region. Wei \emph{et al.} \cite{arXiv_2022_Wei_Tri} investigated the utilization of triple polarization for multi-user HMIMO systems, aiming for increasing both the capacity and diversity without enlarging the antenna array dimension. Near-field propagation takes place between the transceiver surfaces, which consist of compact sub-wavelength triple polarization patch antennas. Based on the dyadic Green’s function, a pair of TPC schemes are designed for mitigating cross-polarization and inter-user interference. It is demonstrated that triple polarization HMIMO is capable of increasing the channel capacity by a factor of three compared to its conventional uni-polarization counterparts. In a nutshell, by leveraging near-field propagation and polarization strategies, it is possible to significantly improve the performance of HMIMO systems, particularly in the near-field region.
\begin{figure}[!t]
\centering
\subfloat[]{\includegraphics[width=8cm]{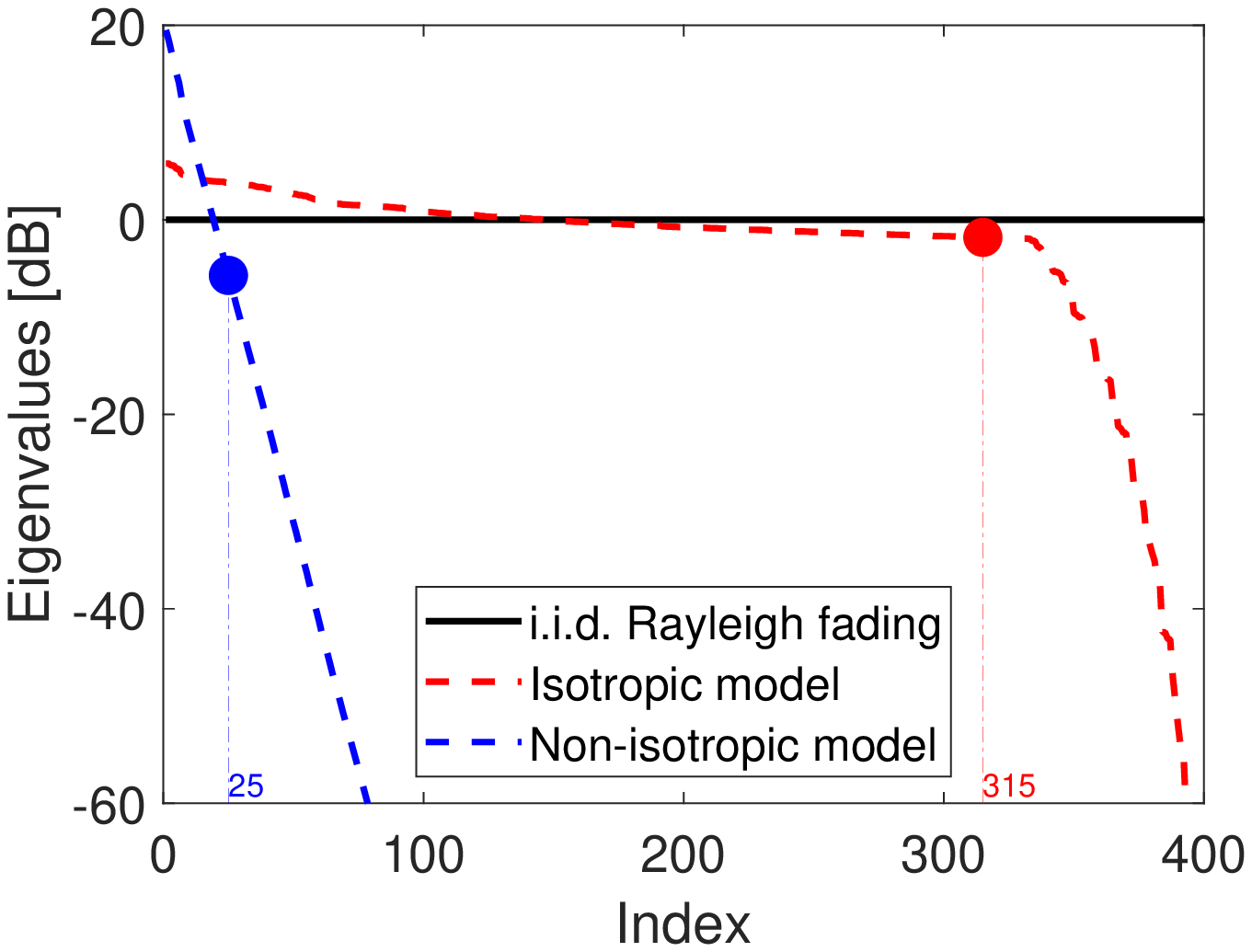}%
\label{fig_2a}}
\subfloat[]{\includegraphics[width=8cm]{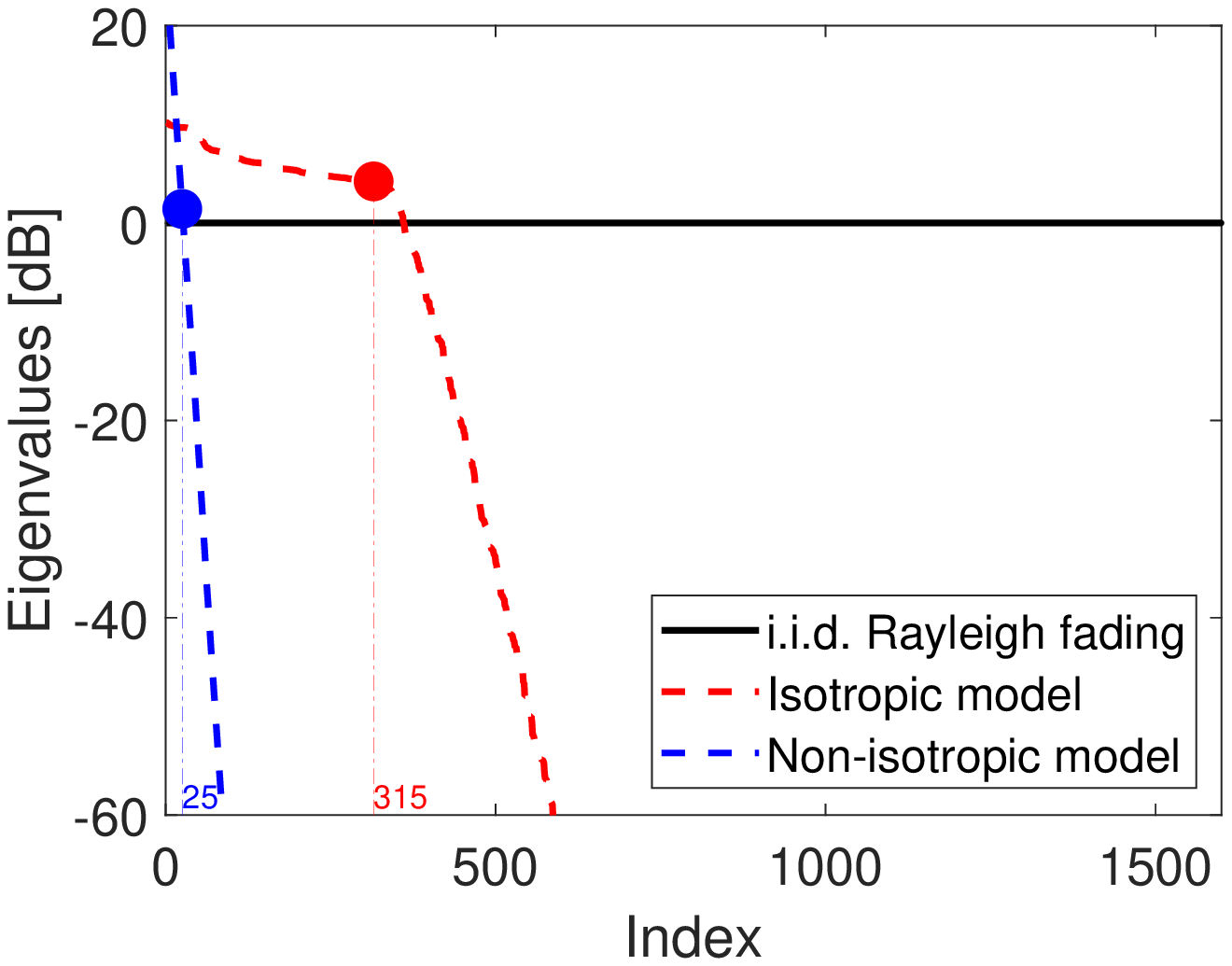}%
\label{fig_2b}}
\caption{Channel eigenvalues of $\mathbf{R}_{r}$ sorted in descending order for a squared array of size $10\lambda \times 10\lambda $: (a) array with $\lambda /2$ antenna spacing ($N_{r} = 400$); (b) array with $\lambda /4$ antenna spacing ($N_{r} = 1600$).}
\label{fig_2}
\end{figure}
\begin{figure}[!t]
\centering
\subfloat[]{\includegraphics[width=8cm]{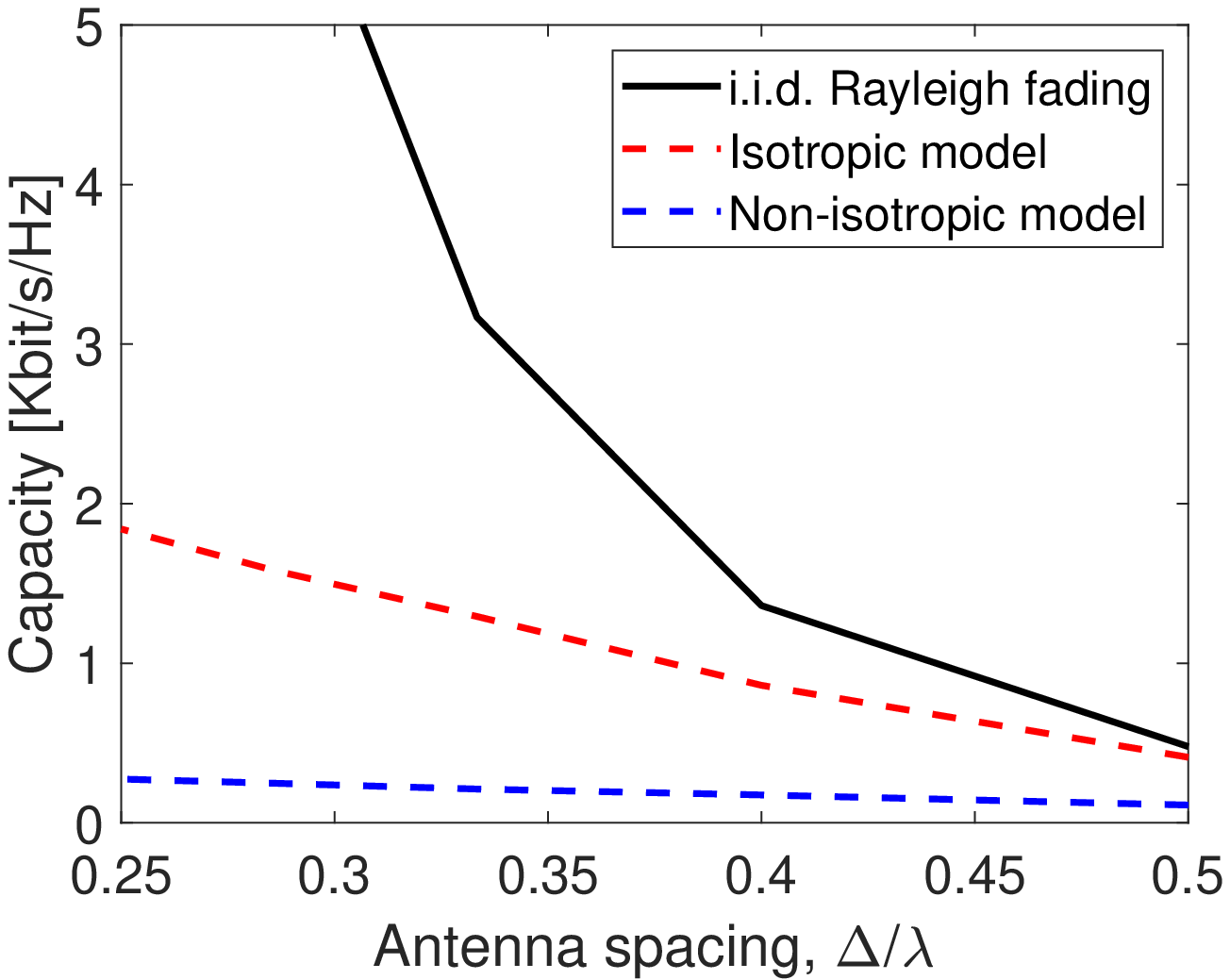}%
\label{fig_3a}}
\subfloat[]{\includegraphics[width=8cm]{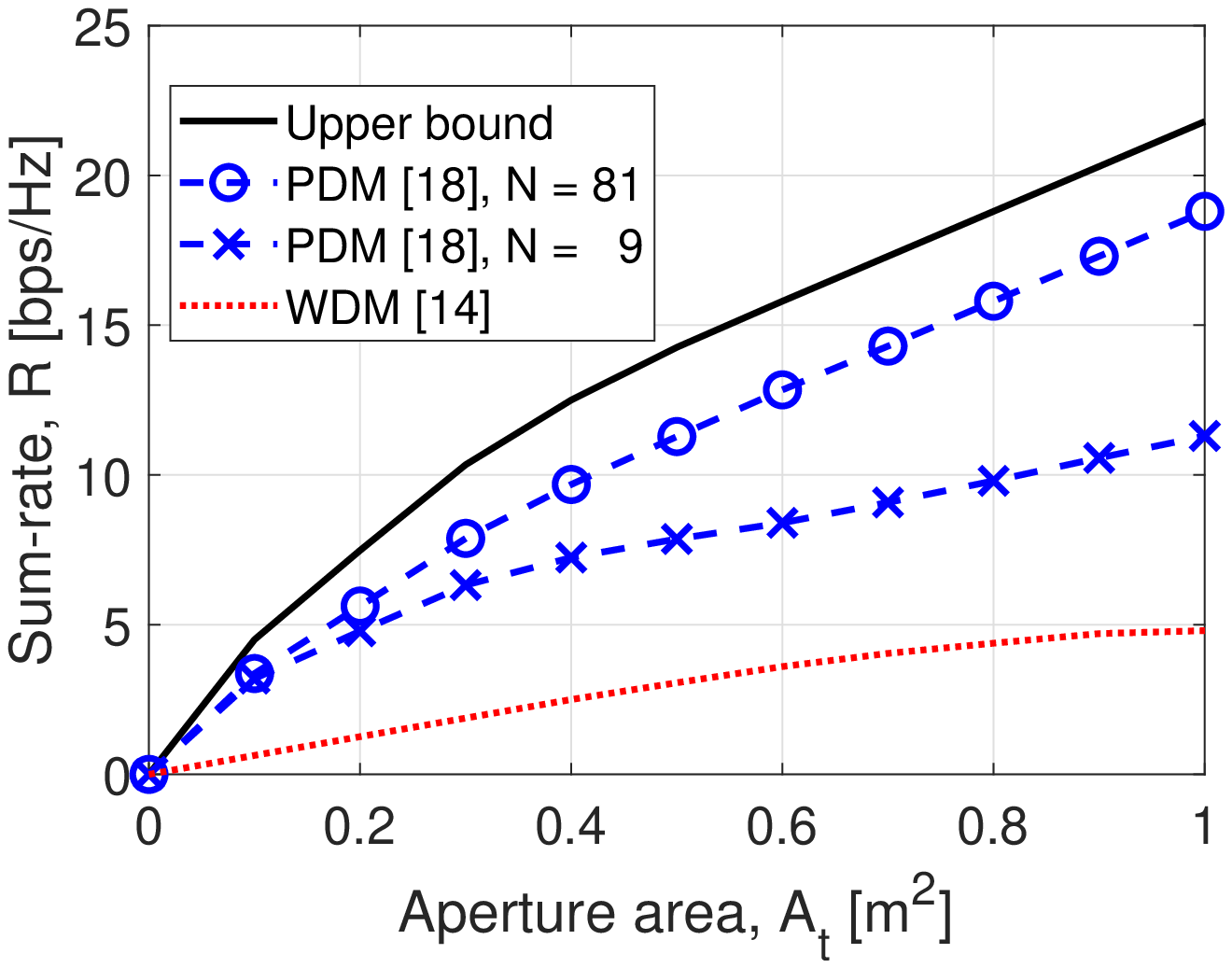}%
\label{fig_3b}}
\caption{(a) Ergodic capacity $C$ in Kbit/s/Hz versus antenna spacing $\Delta \in \left [ \lambda /4,\lambda /2 \right ] $ for a squared array of size $L = 10$ and $\gamma =10$ dB; (b) Sum-rate $R$ versus the aperture area $A_t$.}
\label{fig_3}
\end{figure}
\section{Numerical Results}\label{sec4}
In this section, we will provide numerical results for quantifying the spatial DoF and channel capacity of an HMIMO system.
\subsection{Spatial DoF Evaluation}

The correlation properties of $\mathbf{H}$ depend on the eigenvalues of $\mathbf{R}$. Specifically, for the Kronecker model of \eqref{eq3} associated with symmetric scattering, we only have to consider the eigenvalues of $\mathbf{R}_{r}$. Fig. \ref{fig_2} displays the sorted eigenvalues for $L_{r,x}=L_{r,y}=10$, where we consider two setups, one using $\lambda /2$- and another having $\lambda /4$-spaced antenna elements, which correspond to $N_{r}=400$ and $N_{r}=1600$, respectively. Both isotropic and non-isotropic propagation conditions are considered. The number of significant coupling coefficients for these two cases is $n_{r}'=n_{r}\approx 315$ and $n_{r}'=25$, respectively, which are indicated as a dot on the corresponding curves. These values determine the number of eigenvalues that carry the fundamental channel information. Note that in the i.i.d. Rayleigh fading model, all $N_{r}N_{t}$ eigenvalues are equal to $1$. Hence, the gap between the two models is determined by $N_{r}/n_{r}=\lambda ^{2}/\left ( \pi \Delta _{r,x} \Delta _{r,y} \right )$, assuming uniform antenna spacings of $\Delta _{r,x},\Delta _{r,y}$. This gap yields a roughly $1.2\times $ and $5\times $ overall increase in the number of eigenvalues, which grows quadratically with the normalized antenna spacing.

Moreover, it can be observed from Fig. \ref{fig_2} that as the coupling coefficients become more uneven, the eigenvalues decay more steeply, resulting in a higher degree of correlation. Ideally, if $\mathbf{\Lambda }=N_{t}N_{r}\mathbf{I}_{n_{t}n_{r}}$, the channel samples would be mutually independent. Additionally, observe from Fig. \ref{fig_2} that the HMIMO model of an isotropic scattering environment exhibits an $n_{r}$-order low-rank property. The error caused by discarding the remaining $\left ( N_{r}-n_{r} \right )$ eigenvalues is approximately $4.6$\% of the total channel power at $L_{r,x}=L_{r,y}=10$ and tends to zero asymptotically as the element spacing is further reduced.

\subsection{HMIMO Performance Evaluation}
In Fig. \ref{fig_3a}, we plot the ergodic capacity in Kbit/s/Hz versus antenna spacing $\Delta$, using the same setup as illustrated in Fig. \ref{fig_2a} with $L =10$. Note that under Clarke’s isotropic model, the power is allocated only to the most significant $n_{t}$ eigenmodes. Compared to the i.i.d. Rayleigh fading, a large capacity gap is observed for $\Delta < \lambda /2$ due to the correlation that naturally arises among antennas as $\Delta$ decreases. The capacity may be further degraded under the non-isotropic scenario, as depicted in Fig. \ref{fig_3a}.

Fig. \ref{fig_3b} plots the sum-rate $R$ versus the aperture area $A_t$, where we consider PDM and WDM beamforming techniques. The specific experimental setups are the same as in \cite{arXiv_2021_Zhang_Continuous}. Note that the PDM achieves a higher sum-rate due to the enhanced orthogonality among the patterns associated with $K$ users. In addition, the PDM scheme achieves performance improvement as the number of elements $N$ increases. Since a small number of elements cannot fully utilize the spatial DoF of practical HMIMO channels, it is crucial to deploy an appropriate number of elements to approximate the continuous array.
\section{Conclusions}\label{sec5}
In this letter, we have critically appraised the performance analysis of the HMIMO communications and surveyed the recent progress in holographic beamforming. We began by introducing the spatial DoF based on the statistical properties of the HMIMO channel. It is demonstrated that the full DoF of HMIMO is determined by the smallest area of the transmitting and receiving surfaces, normalized to the squared wavelength. We then revisited the recent advances in the performance analysis of HMIMO systems. Moreover, we highlighted the recent advances in holographic beamforming and beam-focusing in various HMIMO scenarios. Numerical results have been provided for evaluating both the spatial DoF and the HMIMO capacity.
\bibliographystyle{IEEEtran}
\bibliography{ref}
\end{document}